\documentclass[runningheads,a4paper]{llncs}
\usepackage{graphicx}
\usepackage{mathrsfs}
\usepackage{latexsym}
\usepackage{amssymb}
\usepackage{amscd}
\usepackage{amsmath}
\usepackage{bbm}
\usepackage{url}
\usepackage{float}
\usepackage{changepage}
\usepackage{xcolor}
\usepackage{booktabs}
\usepackage{enumerate} 

\newcommand{\mli}[1]{\mathit{#1}}

\newcommand{\keywords}[1]{\par\addvspace\baselineskip
\noindent\keywordname\enspace\ignorespaces#1}

\begin{document}

\mainmatter 

\title{Episodic source memory over distribution by quantum-like dynamics -- a model exploration.}

\titlerunning{Hamiltonian-QEM}


\author{ Broekaert, J.,  Busemeyer, J. }
%
%
\authorrunning{ Broekaert \& Busemeyer} 
%
\tocauthor{ Jan Broekaert }
\institute{Department of Psychological and Brain Sciences, \\ Indiana University, Bloomington. \\
\email{jbbroeka@indiana.edu, jbusemey@indiana.edu}
}

\maketitle

\begin{abstract}
In \emph{source memory} studies, a decision-maker is concerned with identifying the context in which a given episodic experience occurred. A common paradigm for studying source memory is the `three-list' experimental paradigm, where a subject studies three lists of words and is later asked whether a given word appeared on one or more of the studied lists. Surprisingly, the sum total of the acceptance probabilities generated by asking for the source of a word separately for each list (`list 1?', `list 2?', `list 3?') exceeds the acceptance probability generated by asking whether that word occurred on the union of the lists (`list 1 or 2 or 3?'). The episodic memory for a given word therefore appears \emph{over distributed} on the disjoint contexts of the lists. A quantum episodic memory model [QEM] was proposed by Brainerd, Wang and Reyna (2013) to explain this type of result. In this paper, we apply a Hamiltonian dynamical extension of QEM for over distribution of source memory. 
The Hamiltonian operators  are simultaneously driven by parameters  for  re-allocation of  \emph{gist}-based and \emph{verbatim}-based acceptance support as subjects are exposed to the cue word in the first temporal stage, and  are attenuated for description-dependence by the querying probe in the second temporal stage. 
 Overall, the model predicts well the choice proportions in both separate list and union list queries and the over distribution effect, suggesting that a Hamiltonian dynamics for QEM can provide a good account of the acceptance processes involved in episodic memory tasks.
\end{abstract}

\section{Familiarity and recollection, verbatim and gist \label{introduction}}
Recognition  memory  models predict  judgments  of  `prior  occurrence  of  an  event'.  In recognition, Mandler  distinguished  a \emph{familiarity} process  and  a \emph{retrieval} - or \emph{recollection} - process  that would  evolve  separately but also additively \cite{mandler1980}.  
The familiarity of a memory  would  relate to  an `intra event  organizational   integrative  process', while retrieval relates to an  `inter event  elaborative  process'.  
 Extending this \emph{dual process} modeling work, by Tulving \cite{tulving1984} and  Jacoby \cite{jacoby1991}, a `conjoint recognition' model was developed by Brainerd, Reyna and Mojardin \cite{brainerdetal1999}  which provides separate parameters for the entangled processes of identity judgement, similarity judgment and response bias.
Their model implements \emph{verbatim} and \emph{gist} dimensions to memories. Verbatim traces hold the detailed contextual features of a past event, while gist traces hold its semantic details. In recognition tasks we would access  \emph{verbatim} and \emph{gist} trace in parallel.  The verbatim trace of a verbal cue handles it surface content like orthography and phonology for words with its contextual features like in this case, colour of back ground and text font. The verbal cue's gist trace will encode relational content like semantic content for words, also with its contextual features.  This development recently received a quantum formalisation for its property of superposed states to cope with \emph{over distribution} in memory tests \cite{busemeyerbruza2012,brainerdetal2013,brainerdetal2015}. 
In specifically designed expermental tests it appeared episodic memory of a given word  is \emph{over distributed} on the disjoint contexts of the lists, letting the acceptance probability behave as a \emph{subadditive} function \cite{brainerdreyna2008,brainerdetal2015}. \\
{\bf Quantum-like memory models}   
The  Quantum Episodic Memory model (QEM) was proposed by Brainerd, Wang and Reyna (2013). It assumes a Hilbert space representation in which verbatim, gist, and non-related components are orthogonal, and in which recognition engages the gist trace in target memories as well. We will provide ample detail about this model in the next section, since our dynamical extension is implemented in essentially the same structural setting.
QEM was extended  to generalized-QEM (GQEM) by Trueblood and Hemmer  to model for \emph{incompatible} features of gist, verbatim and non-related traces  \cite{truebloodhemmer2017}. Subjacent is the idea that these features are serially processed, and that gist precedes verbatim since it is processed faster.
Independently, Denolf and Lambert-Mogiliansky have considered the accessing of gist and verbatim as incompatible process features.  This aspect  is implemented in an intrinsically quantum-like manner in their \emph{complementarity}-based model for Complementary Memory Types  (CMT) \cite{denolf2015,denolflambert2016,lambertmogiliansky2014}. 
We previously developed a Hamiltonian dynamical extension of QEM for \emph{item} memory  tasks \cite{broekaertbusemeyer2017}. The dynamical formalism allows to describe time development of the acceptance decision based on gist, verbatim and non-related traces. 
Finally, also a semantic network approach by Bruza, Kitto,  Nelson, and McEvoy \cite{nelsonetal2013} was developed in which the target word is adjacent to its associated terms and the network is in a quantum superposition state of either complete activation or non-activation (see also \cite{bruzaetal2009}).  \\
We note that dynamical approaches to quantum-like models have been proposed previously, e.g. in decision theory by Busemeyer and Bruza \cite{busemeyerbruza2012}, Pothos and Busemeyer \cite{pothosetal2013},  Mart\'inez-Mart\'inez \cite{martinez2014} and by Yearsley and Pothos \cite{yearsleypothos2016},  in cognition by  Aerts, Broekaert and Smets \cite{aertsetal1999}, in perception theory by Atmanspacher and Filk \cite{atmanspacherfilk2010}. 

\section{True memory, false memory, over distributed memory.}
\begin{figure}[h]
 \begin{minipage}[b]{.5\linewidth}
   \centering
  \begin{tabular}[b]{@{} cccl @{}}
    \toprule
$L_i?$& $L_j?$ & $L_k?$ & \ $\vert$ \ \ $L_i$ \\ 
    \midrule
   yes 	& yes 	& yes	& $\to$ \ over distribution \\ 
   yes 	& yes 	& no 		& $\to$ \ over distribution \\ 
   yes 	& no 		&yes 	& $\to$ \ over distribution\\ 
   yes 	& no 		& no 		& $\to$ \ true memory \\ 
    no 	& yes 	&yes 	& $\to$ \ false memory \\ 
    no 	& yes 	& no 		& $\to$ \ false memory \\ 
    no 	& no 		& yes 	& $\to$ \ false memory \\ 
    no 	& no 		& no 		& $\to$ \ forgotten \\ 
       \bottomrule
 \end{tabular}  
\end{minipage}
  \begin{minipage}[b]{0.50\linewidth}
    \centering
\includegraphics[width=\linewidth]{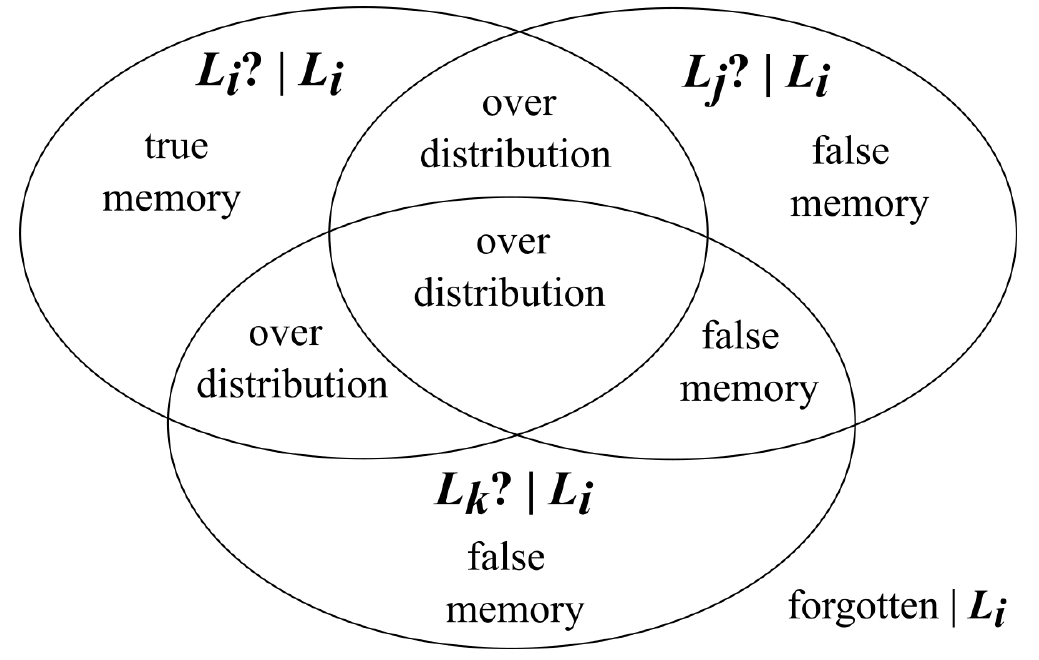}
  \end{minipage}%
 \caption{Logic of false memory, true memory  and over distribution in the three list paradigm for source memory, for a target which is a studied word from list $L_i$. Indices $[i,j,k]$  are permutations of [1,2,3].  For a distractor, which is an unstudied word from $L_4$, all response triplets are erroneous memories, except the triple `no'  which is a correct no-memory evaluation. }
\label{figtable}
\end{figure}

In the \emph{conjoint process dissociation} model (CPD) a sufficient parametrisation is present to capture the four distinct response patterns of true, false, over distributed and forgotten memories of the three-list  paradigm.  The precise identification of the types of memories for a given target requires a composite outcome for the acceptance to three lists at once (see fig. and table \ref{figtable} ).  For instance, should a participant report ``the word appeared on $L_1$, on $L_2$ but not on $L_3$," when the cue word came from list $L_2$ then this participant clearly showed a case of memory over distribution.  If however that same answer had been given for a cue coming from  list $L_3$ then this participant  showed a case of false memory. \\
The participant is however well informed at the start that the word lists do not overlap. It makes therefor no sense  to ask for an answer to a conjunctive composition query probe at one instance: multiple-yes answers would be absent and therefor no cases of over distribution could be produced. A quantum based model for the  \emph{conjunction} of queries  moreover requires a procedure specification for its formal representation, since measurement outcomes  in quantum models  are  sensitive to ordering of the measurement operators for non-compatible questions \cite{aerts2009,wangbusemeyer2013,pothosetal2013,busemeyerbruza2012}.  While the projectors for list membership, Eqs. \ref{projectors},  are commutative the dynamical process between two measurements will void that order invariance, as we will see in the next section.
The dynamical process implies that the Hamiltonian-QEM predicts  different acceptance probabilities for different query orderings, e.g. $p (L_i  ?  \circ  L_j  ? \vert L_i ) \not = p (L_j  ?  \circ  L_i ? \vert L_i )$. It is therefor not possible in Hamiltonian-QEM to define a unique expression for expressions like $p (\neg L_i ? \cap L_j ? \cap  L_k  ? \vert L_i )$ (cfr \cite{brainerdetal2013}) without additional information on the order of querying.

In the three list experimental paradigm the query probes are kept separated -- `did the word appear on List 1' ($L_1?$), `did the word appear on List 2' ($L_2?$), `did the word appear on List 3' ($L_2?$) and the disjunctive probe  `did the word appear on one of the lists' ($L_{123}?$) -- and  are randomised between other acceptance  tasks  for other words \cite{brainerdetal2013,truebloodhemmer2017,denolflambert2016,kellenetal2014}. \\
From a classical set theoretic perspective we can relate the acceptance probability for the disjunctive probe with the acceptance probabilities of the single probes;
\begin{eqnarray}
\vert S( L_{1} \cup L_{2} \cup L_{3}?) \vert  &=& \textstyle  \sum_i\vert S( L_{i} ?) \vert    - \textstyle\sum_{i<j} \vert S( L_{i} \cap L_{j} ?)\vert   + \vert S( L_{1} \cap L_{2} \cap L_{3}?) \vert \ \ \ \ \ \  \label{setrelation}
\end{eqnarray}
(for a word from a given set $L_k$). Then we define the unpacking factor $\mli{UF}$ as the ratio of  number of acceptance responses for the separate queries per list over the number of cases of the query for the joined lists : 
\begin{eqnarray}
 \mli{UF}(k) &=& \frac{\textstyle  \sum_i\vert S( L_{i}) \vert}{  \vert S( L_{1} \cup L_{2} \cup L_{3}?) \vert }
\end{eqnarray}
In terms  of summed acceptance probabilities and taking into account classical set relation, Eq. (\ref{setrelation}), and some algebra, the interpretation of the unpacking factor is apparent;
 \begin{eqnarray} 
 \mli{UF}(k)  &=& \frac{p(L_1? \vert L_k)  + p(L_2? \vert L_k) + p(L_3?\vert L_k) }{p(L_{123}?\vert L_k)} \nonumber \\
    &=& 1 + \frac{p(L_1 \cap L_2 \cap \neg L_3 ?\vert L_k)  + p(\neg L_{1} \cap L_{2} \cap L_{3}?\vert L_k) + p(L_{1 } \cap \neg L_{2} \cap L_{3}?\vert L_k) }{p(L_{123}?\vert L_k)}  \nonumber \\
    & & + 2\,  \frac{ p(L_{1 } \cap L_{2} \cap L_3?\vert L_k) }{p(L_{123}?\vert L_k)}  \label{UF}
 \end{eqnarray}
 For every index value $k$ of the target's list, the excess value of $ \mli{UF}$ above 1 is caused by three over distribution terms of which the `always accept' term is double weighted, and one false memory term of the type `accept on all lists except the true source'.  For example, when $k=1$, the term $\frac{p(\neg L_{1} \cap L_{2} \cap L_{3}?\vert L_1) }{p(L_{123}?\vert L_1)} $ relates to the case a target from $L_1$ was not accepted on that list while it was accepted both on $L_2$ and $L_3$, constituting a false memory contribution to $ \mli{UF}$.\\
Since the lists in the experimental design are disjoint, according to classical logic the right hand side is equal to 1. For experimental choice proportions however the unpacking factor shows to be significantly larger than 1 \cite{brainerdetal2013}. Modulo the fact that along three terms for over distribution the unpacking factor always mixes in one term of false memory as well,  we will still use the unpacking factor as a measure for over distribution, besides its correct measure for subadditivity.\footnote{In principle contributions of false memories and over distributions could be fully separated if one would measure the acceptance probabilities for disjunctions of all disjunctive list \emph{pairs} as well. }

\section{The Hamiltonian based QEM model.}
In essence the  Hamiltonian based QEM model describes in two subsequent temporal stages how the belief state of the participants evolves through the experimental paradigm. This change of the belief state is described by two distinct Schr\"odinger evolutions. In the first stage the participant is presented with a cue originating from one of the four lists - three of them with targets, one with distractors. First, the participant processes this incoming information to change her initial `uniform' belief state into a state informed by the presented cue and her memory. Subsequently this state of belief is then further evolved due to the processing of the information of the probe. The information of the probe allows for a response bias or description-dependence. We expect the latter evolution to be an attenuation of the first stage recognition phase.\\
{\bf State vectors.} In line with Brainerd, Wang and Reyna's development of QEM,  the state vector is expressed on the orthogonal basis ($V_1,V_2,V_3,G,N$). The model thus provides a dedicated dimension for verbatim support for each list, and a dimension for gist support shared for all lists. The last dimension is dedicated to support for non-related items or distractors. In our dynamical development of QEM each state vector is modulated according the cue and probe combination to which the subject is exposed
\[ \Psi_{ probe  \vert cue  }(t) =   [{\psi_{p \vert c }}_{V_1}(t) ,{\psi_{p \vert c }}_{V_2}(t) ,{\psi_{p \vert c }}_{V_3}(t) ,{\psi_{p \vert c }}_G (t), {\psi_{p \vert c  }}_N (t)]^{\tau}, \] 
amounting to sixteen distinct states in the present experimental paradigm.\\

{\bf The initial state vector.}  In the basis ($V_1,V_2,V_3,G,N$), the generalized initial state vector is expressed as
\begin{eqnarray}
\Psi_0 (g) = \left[ \sqrt{(1-g^2)/6},  \sqrt{(1-g^2)/6},  \sqrt{(1-g^2)/6} , g/\sqrt{2},  1/\sqrt{2} \right]^\tau \label{initialstate}
\end{eqnarray}
where we restrict the parameter $g \in [-1,1]$.  This initial belief state of the subject reflects to certain extent the fact - of which the subject is informed -  that in this experiment  half of the cues are non-studied, and half of them come from the three studied lists (p. 419, \cite{brainerdetal2012}).  This form also implements the idea that the subject at the start has a latent tendency for acceptance of the cues.  This form of the initial state expresses that a cue from a studied list elicits on average  $(1-g^2 )/6$ acceptance probability from verbatim trace and $g^2/2$ acceptance probability from the gist trace.\footnote{Without taking into account of dynamics for the effect of cue or probe, but still applying the measurement projections  Eqs (\ref{projectors}),   the amount of gist  $g$ in the initial state shows a latent tendency for overdistribution
\begin{eqnarray}
p_0(L1)+p_0(L2)+p_0(L3)+p_0(N) = 1+g^2 \nonumber
\end{eqnarray} 
with $p_0(N) =1/2$. Clearly the experimental description `half of the cues are $N$, the other half originate from the lists', cannot be implemented exactely due to overdistribution.}
The  parameter $g$ therefor indicates the preponderance of gist in the initial state for a given cue.  It should be noticed however that the initial state itself is not measured on. In the Hamiltonian model a dynamical evolution transforms the initial belief state till the point of measurement. Our present assessment of the initial component amplitudes therefore only concern a latent tendency. We will at present fix the initial state to correspond mathematically to equal weighting of acceptance support by verbatim trace  and by  gist trace for the three studied lists.  The initial state then reads, with $g=0.5$  in Eq.(\ref{initialstate}), 
\begin{eqnarray}
\Psi_0 = \left[ 1/2\sqrt{2},1/2\sqrt{2},1/2\sqrt{2},1/2\sqrt{2},  1/\sqrt{2} \right]^\tau \label{initialstate}
\end{eqnarray}

\begin{figure}[h]
   \includegraphics[width=5in]{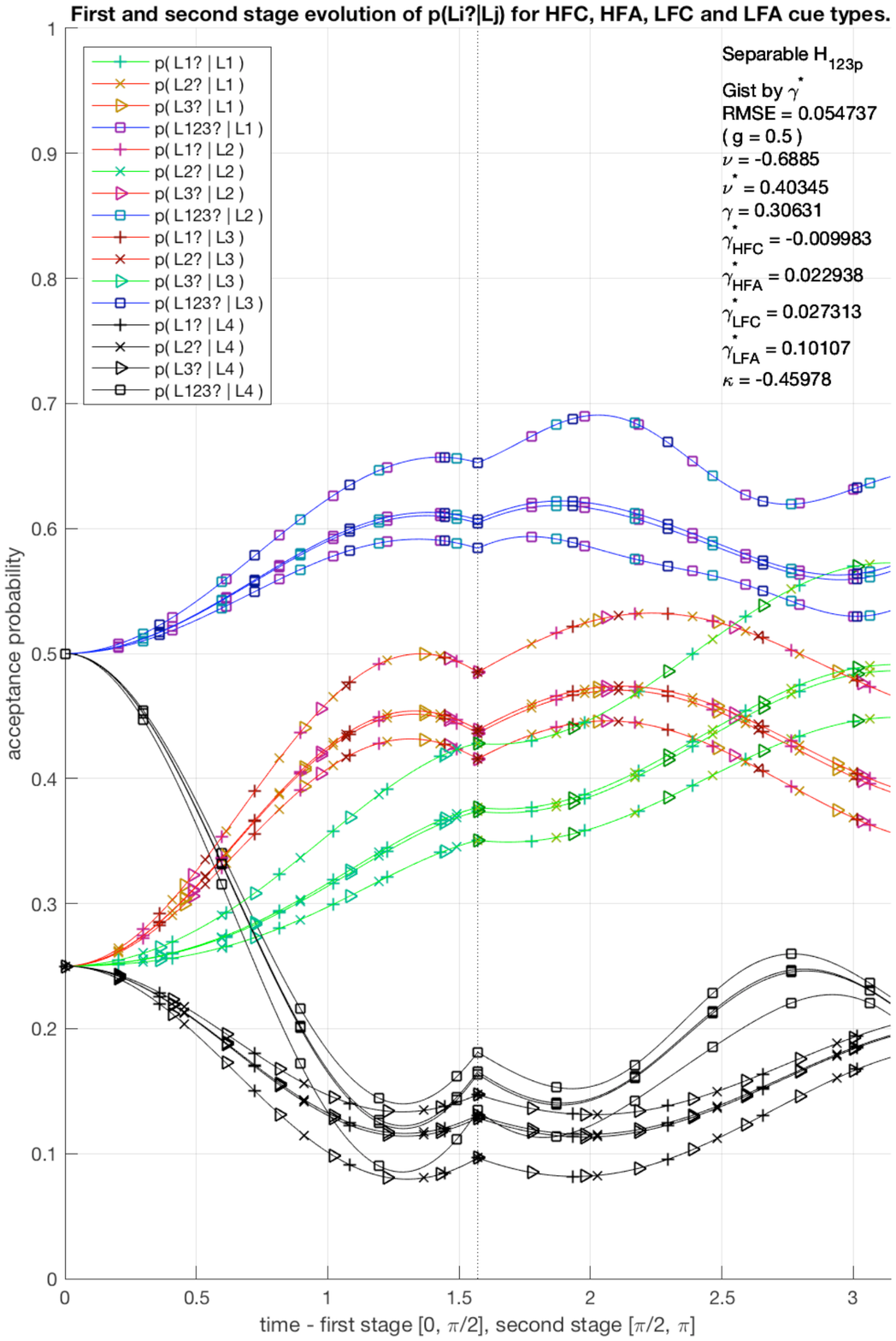} 
   \caption{Optimized two-stage temporal evolution of the acceptance probabilities $p(L_i? \vert L_j)$ according the Hamiltonian-QEM model with separable $H_{123p}$ for querying the joint-list, and with equal latent gist and verbatim support in the initial state ($g=0.5$).}
   \label{evolutionfigure}
\end{figure}

{\bf Measurement projectors}
The subject's response to the probes $L_i?$, or \emph{$L_{123}?$} for a given cue of  $L_j$, are obtained by applying the measurement operators on the final state $\Psi_{ probe  \vert cue}$. The measurement operators are projector matrices which select the components of the final outcome vectors for the specific response. The projector for e.g. $L_1?$ must select both the dedicated verbatim amplitude ${\psi_{L_1 \vert c }}_{V_1}$ and the gist component ${\psi_{L_1 \vert c }}_G$. In ($V_1,V_2,V_3,G,N$) ordered Hilbert space the corresponding projector $M_{L_1?}$ will thus collapse the outcome state to a subspace spanned  on the $(V_1,G)$ basis, and is implemented by a matrix with diagonal elements (1,0,0,1,0) and 0 otherwise. All projectors for the three list paradigm are implemented accordingly:
\begin{eqnarray} 
 \scriptstyle
M_\mli{L_1?} = \left[ \begin{array}{ccccc} 1 & 0 & 0 & 0 & 0 \\ 0 & 0 & 0 & 0 & 0 \\ 0 & 0 & 0 & 0 & 0  \\ 0 &0 &0 & 1& 0 \\ 0 & 0 & 0 & 0 & 0 \end{array}\right], \, 
M_\mli{L_2?} = \left[ \begin{array}{ccccc}  0 & 0 & 0 & 0 & 0 \\ 0 & 1 & 0 & 0 & 0 \\ 0 & 0 & 0 & 0 & 0  \\ 0 &0 &0 & 1& 0 \\ 0 & 0 & 0 & 0 & 0 \end{array}\right], \,
M_\mli{L_3?} = \left[ \begin{array}{ccccc}  0 & 0 & 0 & 0 & 0 \\ 0 & 0 & 0 & 0 & 0 \\ 0 & 0 & 1 & 0 & 0 \\ 0 &0 &0 & 1& 0 \\ 0 & 0 & 0 & 0 & 0 \end{array}\right], \, 
M_\mli{L_{123}?} = \left[ \begin{array}{ccccc}  1 & 0 & 0 & 0 & 0 \\ 0 & 1 & 0 & 0 & 0 \\ 0 & 0 & 1 & 0 & 0 \\ 0 &0 &0 & 1& 0 \\ 0 & 0 & 0 & 0 & 0 \end{array}\right] \ \ \ \ \label{projectors} 
\end{eqnarray}
We notice the acceptance projectors commute, but are not orthogonal as they all include the gist component in their projective subspace;
 \begin{eqnarray}    
 \left[ P_\mli{L_i?}, P_\mli{L_j?} \right] = 0 , & &    P_\mli{L_i?}. P_\mli{L_j?} \neq 0 \ \ \ \ \ \ \   (i\neq j) \nonumber
 \end{eqnarray}
Trueblood and Hemmer \cite{truebloodhemmer2017}, and Denolf and Lambert-Mogliansky \cite{denolflambert2016} point out QEM's orthogonality of verbatim, gist and non-related features need not necessarily be retained (cfr model specifications in Section \ref{introduction}). In our dynamical extension of QEM a query probe engenders its proper dynamics, therefor repeated application of projectors, without intermediate evolution, will not occur. \\
{\bf Acceptance probabilities.}  
In quantum-like models probabilities are given by the squared length of the projected outcome vectors.
Using the projector operators, eqs (\ref{projectors}), the resulting  acceptance probabilities  for a given probe $L_i?$ after a given cue $L_j$, are explicitly given by the expressions:
\begin{align}
p(L_i?\vert L_j) &= \scriptstyle \left\vert  {\psi_{L_i?\vert L_j}}_{V_i} \right\vert^2  + \left\vert {\psi_{L_i?\vert L_j}}_G \right\vert^2,    \label{pLij}\\
p(L_{123}?\vert L_j) &= \scriptstyle \left\vert {\psi_{L_{123}?\vert L_j}}_{V_1} \right\vert^2  + \left\vert {\psi_{L_{123}?\vert L_j}}_{V_2} \right\vert^2  +\left\vert {\psi_{L_{123}?\vert L_j}}_{V_3} \right\vert^2  +\left\vert {\psi_{L_{123}?\vert L_j}}_G\right\vert^2    \label{pL123}
\end{align}
Notice that the probe index $i$ runs from 1 to 3, while the cue index $j$ runs from 1 to 4 since it includes the non-studied list $L_4$. \\
With the explicit expressions of the acceptance probabilities we obtain, using some algebra, the potential for super-additivity in the Hamiltonian-QEM model by means of the unpacking factor, for a given cue $c$;
\begin{eqnarray}
\mli{UF} (c) & = &  1 + \scriptstyle \frac{ \left\vert {\psi_{L_{1}?\vert c}}_{V_1} \right\vert^2 - \left\vert {\psi_{L_{123}?\vert c}}_{V_1} \right\vert^2 + \left\vert {\psi_{L_{1}?\vert c}}_{V_2} \right\vert^2 - \left\vert {\psi_{L_{123}?\vert c}}_{V_2} \right\vert^2 + \left\vert {\psi_{L_{1}?\vert c}}_{V_3} \right\vert^2 - \left\vert {\psi_{L_{123}?\vert c}}_{V_3} \right\vert^2}{\left\vert {\psi_{L_{123}?\vert c}}_{V_1} \right\vert^2  +\left\vert {\psi_{L_{123}?\vert c}}_{V_2} \right\vert^2  +\left\vert {\psi_{L_{123}?\vert c}}_{V_3} \right\vert^2  +\left\vert {\psi_{L_{123}?\vert c}}_G \right\vert^2} \nonumber \\
  &  & + \scriptstyle   \frac{  \left\vert {\psi_{L_{1}?\vert c}}_{G} \right\vert^2 +  \left\vert {\psi_{L_{2}?\vert c}}_{G} \right\vert^2+  \left\vert {\psi_{L_{3}?\vert c}}_{G} \right\vert^2  - \left\vert {\psi_{L_{123}?\vert c}}_{G} \right\vert^2   }{\left\vert {\psi_{L_{123}?\vert c}}_{V_1} \right\vert^2  +\left\vert {\psi_{L_{123}?\vert c}}_{V_2} \right\vert^2  +\left\vert {\psi_{L_{123}?\vert c}}_{V_3} \right\vert^2  + \left\vert{\psi_{L_{123}?\vert c}}_G \right\vert^2}   \label{HQEMunpacking}
\end{eqnarray}
From which follows that - given the positive gist balance in the second fraction - most often the Hamiltonian-QEM model will predict subadditivity in the  three list paradigm. However, like in the  Hamiltonian QEM model for the  \emph{ item false memory} paradigm of Broekaert \& Busemeyer \cite{broekaertbusemeyer2017},  we find the model can in principle also account for cases that satisfy the additivity of disjoint sub-events, or violate it in super-additive manner.\\
{\bf Hamiltonians.} The dynamical evolution of the state vectors is determined by the Hamiltonian operators.  The Hamiltonian reflects the cognitive processing which is engendered by the information in the word cue - differently for a cue from the studied lists or the unrelated list.
In the present three-list paradigm these operators are constructed along four transports - or re-allocations - between components of the belief state; i) between gist-based component and  non-cue verbatim-based component ($G \leftrightarrow V_{\neg i}$),  ii)  between gist-based component and  non-related component ($G \leftrightarrow N$),  iii) between cue verbatim-based component and  non-cue verbatim-based component ($V_{i} \leftrightarrow V_{\neg i}$), and iv) between cue verbatim-based component and  non-related component ($V_{i} \leftrightarrow N$).  The Hamiltonians are constructed by combining the off-diagonal parametrised Hadamard gates of  each transport \cite{broekaertbusemeyer2017}.
The Hamiltonian parameter $\gamma$ controls the transport of  acceptance probability amplitude from non-cue verbatim-based components towards the gist-based component, or back. Similarly  $\gamma'$ regulates  transport of acceptance probability amplitude of the non-related component from or towards the gist-based component. Precisely the strength $\gamma'$  of this dynamic will be made use of to adjust for the  four distinct types of  word cues `high frequency \& concrete' (HFC), `high frequency \& abstract' (HFA), `low frequency \& concrete' (LFC), `low frequency \& abstract' (LFA). Brainerd and Reyna suggest abstract words have weaker verbatim traces than concrete words and low-frequency words have weaker verbatim traces than high-frequency words \cite{brainerdreyna2005,brainerdetal2012,brainerdetal2013}.  A tendency which suggests  gist based transport  from the non-related component to vary  --- namely $\{\gamma'_{HFC}, \gamma'_{HFA}, \gamma'_{LFC}, \gamma'_{LFA} \}$  --- for these distinct types of  word cues.\\
The parameter $\nu$ controls transport of acceptance probability amplitude between non-cue verbatim-based components and the cue verbatim-based component. Similarly  $\nu'$ controls the transport of the acceptance probability amplitude of the non-related component from or towards the cue verbatim-based component. \\
In our present development we demonstrate the parameters fulfil their intended transport: the parameters $\gamma$ and  $\gamma'$ for gist-based transport, and $\nu$ and  $\nu'$ for  cue verbatim-based transport. 
We could have implemented an effect of \emph{list order} by distinguishing  for each list  the transports of the non-related component ($\gamma'_{i,Abs.type}, \nu'_i$)   for forgetfulness variation per list, or  by distinguishing the transports of the target verbatim-based component ($\gamma_i, \nu_i$)  for gist-verbatim diversified acceptance variation per list.  \\
Diminishing the strength of transports, by quenching the driving parameters,  will let outcome acceptance beliefs tend to concur with the acceptance beliefs inherent to the initial state.\\
{\bf First temporal stage.}    The presentation of the word cue starts the memory process of recollection and familiarity in the subject. For cues from studied lists the hamiltonians $H_{1c}$, $H_{2c}$ and $H_{3c}$ engage verbatim-based and gist-based belief. For non-studied cues from list 4, the hamiltonian $H_{4c}$  engages for the non-related belief.
\begin{eqnarray} 
 H_{1c} (\nu, \nu', \gamma, \gamma') =    \left[  \begin{array}{ccccc}
\scriptstyle	1   &\scriptstyle \nu   		& \scriptstyle   \nu    		& \scriptstyle 0    		&  \scriptstyle     \nu'     \\
\scriptstyle       \nu  &\scriptstyle   -1  		&  \scriptstyle  0     		&  \scriptstyle \gamma   	&  \scriptstyle     0       \\    
 \scriptstyle      \nu  &\scriptstyle  0    		& \scriptstyle  -1     		& \scriptstyle  \gamma    	&  \scriptstyle     0      \\
 \scriptstyle       0    &\scriptstyle \gamma  & \scriptstyle \gamma  	&  \scriptstyle 1    		&  \scriptstyle    \gamma' \\
 \scriptstyle      \nu' &\scriptstyle  0    		&\scriptstyle   0   		&  \scriptstyle  \gamma'   	&  \scriptstyle     -1        
      \end{array}  \right]& ,& \ \ \ 
 H_{2c}  (\nu, \nu', \gamma, \gamma')  =   \left[ \begin{array}{ccccc}
\scriptstyle	-1  & \scriptstyle  \nu  &\scriptstyle   0   & \scriptstyle \gamma    &\scriptstyle  0     \\ 
 \scriptstyle     \nu &\scriptstyle 1     &\scriptstyle  \nu  & \scriptstyle 0    & \scriptstyle  \nu' \\ 
 \scriptstyle        0   &\scriptstyle \nu    &\scriptstyle  -1 &\scriptstyle  \gamma  & \scriptstyle   0      \\
 \scriptstyle      \gamma &\scriptstyle   0   & \scriptstyle   \gamma  &\scriptstyle  1   & \scriptstyle   \gamma' \\
 \scriptstyle        0  &\scriptstyle  \nu' & \scriptstyle  0  &\scriptstyle  \gamma' & \scriptstyle  -1 
      \end{array}   \right], \ \ \  \\
H_{3c} (\nu, \nu', \gamma, \gamma')  = \scriptstyle \left[ \begin{array}{ccccc}
\scriptstyle	-1     & \scriptstyle  0           & \scriptstyle   \nu  &   \scriptstyle  \gamma   & \scriptstyle  0     \\
 \scriptstyle        0     &\scriptstyle -1            & \scriptstyle  \nu    &  \scriptstyle \gamma    &  \scriptstyle 0     \\
 \scriptstyle      \nu    &\scriptstyle \nu          &  \scriptstyle  1     &   \scriptstyle      0          &  \scriptstyle  \nu' \\
 \scriptstyle \gamma  &\scriptstyle  \gamma  & \scriptstyle  0       &  \scriptstyle  1             & \scriptstyle  \gamma'  \\
 \scriptstyle       0      & \scriptstyle   0            & \scriptstyle   \nu'   & \scriptstyle \gamma'  & \scriptstyle  -1  
      \end{array} \right]& ,& \ \ \ 
 H_{4c} ( \nu',  \gamma')  =  \scriptstyle\left[ \begin{array}{ccccc}
\scriptstyle	   -1     &\scriptstyle    0    & \scriptstyle  0    & \scriptstyle      0    & \scriptstyle  \nu' \\
\scriptstyle          0     &  \scriptstyle  -1    &\scriptstyle   0     &  \scriptstyle    0    & \scriptstyle  \nu' \\
\scriptstyle          0      &\scriptstyle    0     &\scriptstyle  -1     & \scriptstyle    0     &\scriptstyle  \nu' \\
\scriptstyle          0      &\scriptstyle    0     &\scriptstyle  0      & \scriptstyle    -1    & \scriptstyle \gamma'  \\
\scriptstyle       \nu' &\scriptstyle \nu' & \scriptstyle\nu' &\scriptstyle    \gamma'   & \scriptstyle  1  
      \end{array} \right] .   \label{hamiltoniansforcues} 
\end{eqnarray}
{\bf Second temporal stage.}
The presentation of the probe starts the subsequent decision process in the subject. The outcome of the cue processing and the subsequent probe can be confirmatory or dissonant.  When probing for a studied list, the Hamiltonian $H_{ip}$  for a probe $L_i?$ is equated to the Hamiltonian $H_{ic}$ for processing a cue from $L_i$, in which the driving parameters are now attenuated by a factor $\kappa$. When probing for the union of lists,  the dedicated Hamiltonian $H_{123p}$  for probe $L_{123}?$ is equated  to the sum of the separate Hamiltonians $H_{ip}$ for processing of studied cues from $L_i$, in which the driving parameters are again attenuated by the factor $\kappa$.  This separable construction of the Hamiltonian reflects the cognitive processing of the union list consists of the parallel processing of membership to the three separate lists, and results in the summed attenuated transport between the non-related component and  equally all verbatim components and the  gist  component.
\begin{eqnarray} 
 H_{1p} (\nu, \nu', \gamma, \gamma', \kappa)       &=&    H_{1c} ( \kappa \nu, \kappa \nu', \kappa \gamma, \kappa \gamma')    \\
 H_{2p}  (\nu, \nu', \gamma, \gamma', \kappa)      &=&    H_{2c} ( \kappa \nu, \kappa \nu', \kappa \gamma, \kappa \gamma')       \\
H_{3p} (\nu, \nu', \gamma, \gamma', \kappa)        &=&    H_{3c} ( \kappa \nu, \kappa \nu', \kappa \gamma, \kappa \gamma')     \\
 H_{123p} (\nu, \nu', \gamma, \gamma', \kappa)   &=&    H_{1p} +  H_{2p} +  H_{3p}     \label{hamiltoniansforprobes} 
\end{eqnarray}
{\bf Parameters.} In recapitulation; the Hamiltonian QEM model uses 5 parameters $\{ \mu, \mu', \gamma, \gamma', \kappa \}$ to describe the dynamics of the subject's belief state to evolve from her prior partially informed expectation to the final decision of acceptance of the cue, accounting for 16 acceptance probabilities $p (L_i? \vert L_j)$.
In Section(\ref{dataANDprediction}), we fitted  the parameters for the four distinct types of frequency and concreteness of the cue words HFC, HFA, LFC and LFA. Between these sets we only changed the gist transport parameter $\gamma'$ ($G \leftrightarrow N$) in the Hamiltonians, respectively $\gamma'_\mli{HFA}$, $\gamma'_\mli{HFA}$, $\gamma'_\mli{LFC}$ and $\gamma'_\mli{LFA}$, and otherwise maintained the same Hamiltonian drivers $\gamma, \nu,\nu'$ and $\kappa$. In total therefor 8 parameters are needed to predict the 64 values of the acceptance probabilities.\\
{\bf Unitary time evolution.} In quantum-like models the Hamiltonian  is the operator for infinitesimal time change of the belief state. The operator $U(t)$ for time propagation over a time range $t$ is given by
\begin{eqnarray} 
U(t) &=& e^{ - i H t}  
\end{eqnarray}
The fully evolved belief state -  a solution of the Schr\"odinger equation - is obtained by applying the unitary time operator $U(t)$ to the initial state.
An inherent  feature with Hamiltonian quantum models is the appearance of oscillations of probability over time. In quantum mechanics, finite dimensional and energetically closed systems are  always periodical. Therefor the initial belief state will re-occur after the proper time period of the system.   In previous work we have argued a third temporal stage, closing the experimental paradigm, should implicitly be supposed in which all driving parameters are set equal to zero \cite{broekaertbusemeyer2017}. The time of measurement $t$  is taken equal to $\frac{\pi}{2}$ for each stage  \cite{busemeyerbruza2012}.\footnote{The Hamiltonian parameters are dependent on the choice of measurement time. }  The  final outcome state vector is thus obtained by concatenating both propagators on the initial state, Eq. (\ref{initialstate}),
\begin{eqnarray} 
\Psi_{p\vert c}  &=& e^{ - i H_{p?}  \frac{\pi}{2} } e^{ - i H_c \frac{\pi}{2} } \Psi_0  
\end{eqnarray}
The processing of cue $c$ in the first stage occurs till $t=\frac{\pi}{2}$, the  processing of probe $p?$ in the second stage takes another time range of $\frac{\pi}{2}$. The time evolution of the acceptance probabilities from initial state processing of cue and processing of probe is shown in figure \ref{evolutionfigure}.

\section{Data and prediction \label{dataANDprediction}}
We use the  experimental 3-list data (N=70), reported in Brainerd, Wang and Reyna  (Table 2, \cite{brainerdetal2013}). The CPD model based `bias-correction' of the acceptance probabilities has been  omitted and appear in Table \ref{dataprediction}. The raw data set was provided by dr. Charles Brainerd.    This same data set is used in the CMT model development by  Denolf and Lambert-Mogiliansky \cite{denolf2015,denolflambert2016}, (their Table1 shows some typos for $L_2$ cues in the HFA, LFC and LFA set).
\begin{table}[htbp]
  \centering 
  \begin{tabular}{@{} lcccccccccrlccccccccc  @{}}
       &   &   HFC & &  &  &  & HFA&   &      & &  &    LFC & &   &   &   & LFA &  &   \\ 
 \cmidrule(r){2-5}   \cmidrule(r){7-10}  \cmidrule(r){12-15}  \cmidrule(r){17-20} 
       {\bf  Obs.}      &$L_1 $    & $L_2 $    &   $ L_3$ 	& $L_4$   &  \  	    & $L_1$ 	& $L_2 $	& $L_3$ 	& $L_4$ &  \                 &$L_1 $         & $L_2 $    &   $ L_3$   & $L_4$   &  \  & $L_1$ 	& $L_2 $   & $L_3$ 	& $L_4$ \\ 
    $L_1?$ 	& 0.52     &  0.31        & 0.30   		&  0.15 	&          & 0.52         & 0.32	&   0.40  	&  0.25 &        	            &   0.59         & 0.31        &  0.34         &  0.11     &   & 0.58             &   0.44     &   0.43     &  0.19 \\ 
    $ L_2?$ 	&  0.33  &  0.35  	&  0.43  		&  0.17 	&   	    &  0.36        &  0.54  	& 0.44  	&  0.24  &    	             &    0.46       & 0.46       & 0.35           &  0.11     &   & 0.61             &  0.53      &  0.58      &  0.21 \\ 
    $L_3?$ 	&  0.38  &  0.35  	&  0.42  		&  0.21 	&   	   & 0.37   	       &   0.38  	&  0.48  	&  0.24  &   	             &    0.41       & 0.34  	& 0.49           & 0.13      &   & 0.53              &  0.38     &  0.52       &  0.17  \\ 
$L_{123}?$	& 0.56  &  0.54  	&  0.60   		& 0.22       & 	   & 0.54   	     &   0.64  	&   0.53 	&   0.26 &  	              &  0.64  	&  0.49 	&  0.56          &  0.13    &   &  0.66              &  0.61     &   0.59      &  0.20\\ 
$ \mli{UF} $    & 2.20 &	1.89 		& 1.91  		&2.45	& 	  & 2.32	     &  1.96		&   2.49	&   2.83 &   	              & 2.31   	& 2.31 	&   2.10         &  2.77    &   & 2.59               &   2.19	& 2.61        &    2.86	\\ 
        {\bf  Pred.} &$L_1 $    & $L_2 $    &   $ L_3$ 	& $L_4$   &  \  	& $L_1$ 	& $L_2 $		& $L_3$ 	& $L_4$ &  \    &$L_1 $    & $L_2 $    &   $ L_3$ 	& $L_4$   &  \  	& $L_1$ 	& $L_2 $		& $L_3$ 	& $L_4$ \\ 
    $L_1?$ 	&  0.45  	&  0.36 	&  0.36   	& 0.20 &   	&  0.49     &0.39        & 0.39        &  0.19   &    &   0.49     &0.39       &0.39    &0.19&  & 0.57    & 0.47     &  0.47  &  0.18  \\ 
   $ L_2?$ 	&   0.36 	&   0.45	&   0.36  	&  0.20&    & 0.39      & 0.49   &  0.39         & 0.19      &   &    0.39     & 0.49     &0.39    &0.19 &    &0.47  &   0.57    & 0.47  &  0.18 \\ 
    $L_3?$ 	&  0.36  	& 0.36  	&  0.45   	& 0.20 &   	& 0.39    & 0.39     & 0.49         & 0.19 &        &     0.39       & 0.39    & 0.49    &0.19 &   & 0.47  &  0.47  &  0.57   &  0.18  \\ 
    $L_{123}?$ 	& 0.53 	&   0.53   	&  0.53 	& 0.23&   & 0.57   & 0.57  & 0.57          & 0.22 &             &  0.57    & 0.57   & 0.57        &0.22 & &   0.64 &  0.64    & 0.64  &  0.21\\ 
  $ \mli{UF} $	&  2.18 	&  2.18 	& 2.18	& 2.70 &	& 2.24 &	2.24	   &2.24 	& 2.65 &	    & 2.24 	     & 2.24  &	2.24	     & 2.64 &    &   2.35  &	2.35 	       & 2.35   & 2.52	\\ 

     \cmidrule(r){2-5}   \cmidrule(r){7-10}  \cmidrule(r){12-15}  \cmidrule(r){17-20} 

  \end{tabular}
  \caption{ \small Observed acceptance ratios and unpacking factors, partitioned by cue type, high frequency \& concrete (HFC), high frequency \& abstract (HFA),
 low frequency \& concrete (LFC), low frequency \& abstract (LFA).  N=70. Data set from Table 2, \cite{brainerdetal2013}. Predicted acceptance probabilities and unpacking factors from the Hamiltonian-QEM model, RMSE= 0.054737.}
  \label{dataprediction}
\end{table}
We optimised the RMSE for 64 data points  and Hamiltonian-QEM predictions using 8 parameters  $\{\mu, \mu', \gamma, \gamma'_\mli{HFC} , \gamma'_\mli{HFA},\gamma'_\mli{LFC}, \gamma'_\mli{LFA}, \kappa\}$.
Using a $3^{8}$  grid in parameter space the best fit produced RMSE=0.054737, for the parameter values in Table \ref{parameters}.\\

\begin{table}[htbp]
  \centering \small
  \begin{tabular}{@{} cccccccc @{}}
  $\nu$	& $ \nu'$	  &  	$\gamma$  &   $\gamma'_\mli{HFC}$  & $\gamma'_\mli{HFA}$	& $\gamma'_\mli{LFC}$ & $\gamma'_\mli{LFA}$ &  $\kappa$	 \\ 
\midrule
 -0.6885   &   0.40345 & 0.30631       &      -0.0099825              &     0.022938                     &    0.027313                &   0.10107                 &  -0.45978    \\
    \midrule
  \end{tabular}
  \caption{Best-fit parameters for Hamiltonian-QEM ($t_1 = t_2=\pi/2$).}
  \label{parameters}
\end{table}

\section{Discussion}
We explored the over distribution predictions of a dynamical extension of Quantum Episodic Memory model by Brainerd, Wang, Reyna and Nakamura (2013, 2015) in the 3-list paradigm. The dynamic processing of the initial belief state in our Hamiltonian-QEM results, over time, in outcome states with adequate values of the acceptance probabilities and the unpacking factor for over distribution of memories. The model further predicts systematic higher acceptance probabilities  of targets to their proper source list, $p(L_i? \vert L_i) >  p(L_j? \vert L_i)$ for $(j\neq i )$, which is curiously not consistently observed in the experimental data set (N=70). The model relies on four types of transport embedded in the Hamiltonians; 
 the parameters $\gamma$ and  $\gamma'$ regulate transport affecting the gist-based  component, and the parameters $\nu$ and  $\nu'$ regulate transport affecting the verbatim-based components. The Hamiltonians of the first stage cue recognition phase receive the attenuation parameter $\kappa$ in the second probe response stage. \\
The Hamiltonian-QEM model succeeds in a qualitatively good fit with 8 dynamical parameters for the 64 data points of the experimental data set Brainerd et al. \cite{brainerdetal2013}. Beyond this exploratory test of the Hamiltonian-QEM for source memory,  the model must still be submitted to a comparative statistical test with recent generalisations and modifications of  the QEM model, in particular the Generalized-QEM model by Trueblood and Hemmer \cite{truebloodhemmer2017} and Complementary Memory Types  model by Denolf and Lambert-Mogiliansky \cite{lambertmogiliansky2014,denolf2015,denolflambert2016}. 

\subsection*{Acknowledgements}
We are grateful to Dr. Charles Brainerd for providing us with the raw data file of the 3-List experiment of \cite{brainerdetal2013}.\\
We thank Dr. Peter Kvam for helpful discussions and comments on parts of this manuscript.
This research was supported by AFOSR (FA9550-12-1-00397).

\end{document}